\documentclass[submission, Proceedings]{SciPost}

\hypersetup{
    colorlinks,
    linkcolor={red!50!black},
    citecolor={blue!50!black},
    urlcolor={blue!80!black}
}

\usepackage[bitstream-charter]{mathdesign}
\DeclareSymbolFont{usualmathcal}{OMS}{cmsy}{m}{n}
\DeclareSymbolFontAlphabet{\mathcal}{usualmathcal}
\makeatletter
\providecommand*{\diff}%
	{\@ifnextchar^{\DIfF}{\DIfF^{}}}
\def\DIfF^#1{%
	\mathop{\mathrm{\mathstrut d}}%
	\nolimits^{#1}\gobblespace}
\def\gobblespace{%
	\futurelet\diffarg\opspace}
\def\opspace{%
	\let\DiffSpace\!%
	\ifx\diffarg(%
		\let\DiffSpace\relax
	\else
		\ifx\diffarg[%
			\let\DiffSpace\relax
		\else
			\ifx\diffarg\{%
				\let\DiffSpace\relax
			\fi\fi\fi\DiffSpace}

\begin{document}

\begin{center}{\Large \textbf{
TMD observables in unpolarised Semi-Inclusive DIS at COMPASS \\
}}\end{center}

% TODO: write the author list here. Use initials + surname format.
% Separate subsequent authors by a comma, omit comma at the end of the list.
% Mark the corresponding author with a superscript *.
\begin{center}
Andrea Moretti \textsuperscript{1$\star$} on behalf of the COMPASS Collaboration
\end{center}

\begin{center}
{\bf 1} University of Trieste and INFN, Trieste Section, Italy
\\

* andrea.moretti@cern.ch
\end{center}

\begin{center}
\today
\end{center}

% For convenience during refereeing (optional),
% you can turn on line numbers by uncommenting the next line:
%\linenumbers

\definecolor{palegray}{gray}{0.95}
\begin{center}
\colorbox{palegray}{
  \begin{tabular}{rr}
  \begin{minipage}{0.1\textwidth}
    \includegraphics[width=22mm]{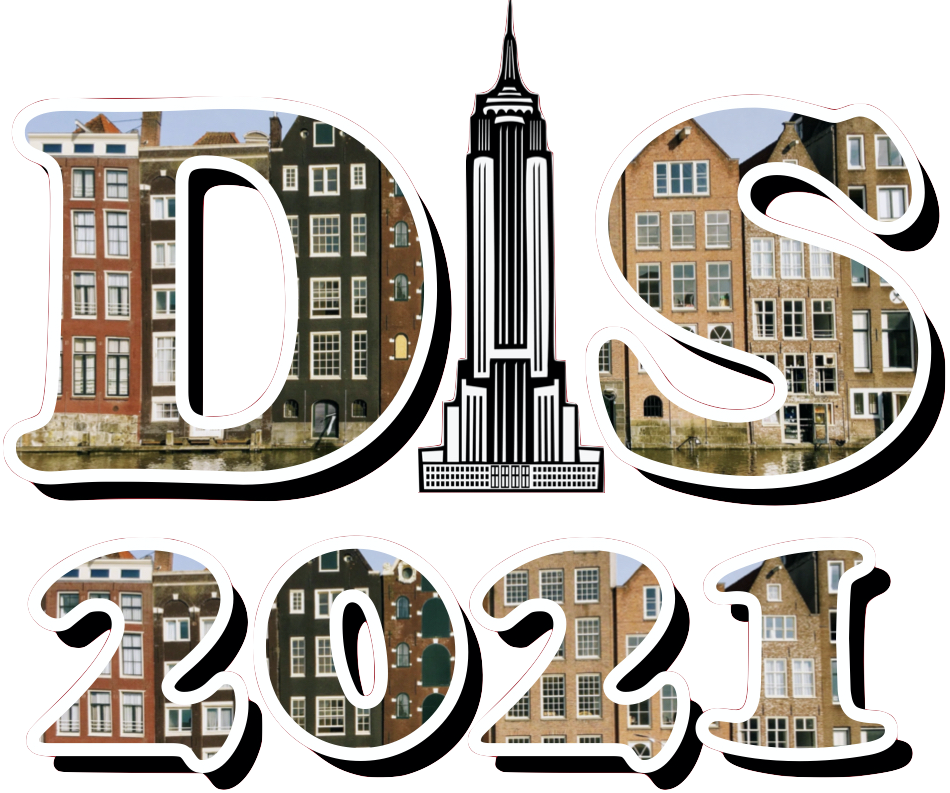}
  \end{minipage}
  &
  \begin{minipage}{0.75\textwidth}
    \begin{center}
    {\it Proceedings for the XXVIII International Workshop\\ on Deep-Inelastic Scattering and
Related Subjects,}\\
    {\it Stony Brook University, New York, USA, 12-16 April 2021} \\
    \doi{10.21468/SciPostPhysProc.?}\\
    \end{center}
  \end{minipage}
\end{tabular}
}
\end{center}

\section*{Abstract}
{\bf
% TODO: write your abstract here.

In 2016 and 2017, the COMPASS Collaboration at CERN collected a large sample of DIS events with a longitudinally polarised 160 GeV/$c$ muon beam scattering off a liquid hydrogen target. Part of the collected data has been analysed to extract preliminary results for the amplitudes of the modulations in the azimuthal angle of the charged hadrons and for their transverse-momentum distributions. These observables give relevant information for the study of the transverse momentum and spin structure of the nucleon. Similarly to the results obtained in COMPASS with a deuteron target, the azimuthal asymmetries exhibit interesting kinematic dependences, while the squared-transverse-momentum distributions can be described by two exponentials.
}

%\vspace{10pt}
%\noindent\rule{\textwidth}{1pt}
%\tableofcontents\thispagestyle{fancy}
%\noindent\rule{\textwidth}{1pt}
%\vspace{10pt}

\section{Introduction}
\label{sec:intro}

The Semi-Inclusive Deep Inelastic Scattering (SIDIS) is a powerful tool to access the nucleon structure.
In the SIDIS process:
\begin{equation}
    \ell(l) + N(P) \rightarrow \ell(l^\prime) + h(P_h) + X(P_X)
\end{equation}
a high energy lepton $\ell$ scatters off a target nucleon $N$ and at least one hadron $h$ is detected in coincidence with the scattered lepton. The quantities in parentheses denote the four-momenta and $X$ represents the unobserved part of the final state. In the Gamma Nucleon System, where the virtual photon and the target nucleon are collinear, the z-axis is along the virtual photon momentum and the x-axis along the leptons transverse momentum, the SIDIS cross-section for an unpolarised nucleon target reads \cite{Bacchetta:2006tn}:

\begin{equation}
    \begin{split}
    & \frac{\diff^5\sigma}{\diff x \diff y \diff z \diff \phi_h \diff P_T^2} = \frac{\alpha^2}{xyQ^2}\frac{y^2}{2\left ( 1-\varepsilon\right )}\left ( 1+\frac{\gamma^2}{2x}\right ) \cdot  \bigg\{ F_{UU,T} + \varepsilon F_{UU,L} \\
    & + \sqrt{2\varepsilon\left ( 1+\varepsilon\right )} F_{UU}^{\cos\phi_h}\cos\phi_h + \varepsilon F_{UU}^{\cos2\phi_h}\cos2\phi_h + \lambda \bigg( \sqrt{2\varepsilon\left ( 1-\varepsilon\right )} F_{LU}^{\sin\phi_h}\sin\phi_h \bigg) \bigg\}.
    \end{split}
\label{eq:sidis_xsec}
\end{equation}
where $x$ is the Bjorken scaling variable; $Q^2$ is the photon virtuality; $y$ is the inelasticity, corresponding, in the laboratory system, to the ratio of the virtual photon energy to the incoming lepton energy; $z$ is the ratio of the energy of the final state hadron to that of the virtual photon as measured in the laboratory; $\phi_h$ is the azimuthal angle of the hadron and $P_T$ is the transverse momentum of the hadron with respect to the virtual photon; $\alpha$ is the electromagnetic coupling constant; $\gamma$ is defined as $\gamma=2Mx/Q$; the kinematic factor $\varepsilon$ corresponds to the ratio of the longitudinal and transverse photon flux; $\lambda$ is the beam lepton longitudinal polarization. 

Five structure functions, indicated as $F_{XY,Z}^{f(\phi_h)}$, where $X$($Y$) indicates the beam (target) polarization, $Z$ the virtual photon polarization and $f(\phi_h)$ is a function of the hadron azimuthal angle, appear in the cross-section. According to the factorization theorem \cite{Collins:2011zzd}, they can be written as convolutions of Transverse-Momentum-Dependent Parton Distribution Functions (TMD-PDFs) and Fragmentation Functions (TMD-FFs). Up to the order $1/Q$ (twist-3):

\begin{equation}
\begin{split}
    F_{UU,T} & =  \mathcal{C}  \left[ f_1D_1 \right] \\
    F_{UU}^{\cos\phi_h} & \simeq \frac{2M}{Q}\mathcal{C} \left[-\frac{(\hat{\textbf{h}}\cdot\textbf{p}_\perp)k_{T}^2}{zM_hM^2}  h_1^\perp H_1^\perp - \frac{\hat{\textbf{h}}\cdot\textbf{k}_T}{M} f_1 D_1\right] \\ 
    F_{UU}^{\cos2\phi_h} & = \mathcal{C} \left[-\frac{2(\hat{\textbf{h}}\cdot\textbf{k}_T)(\hat{\textbf{h}}\cdot\textbf{p}_\perp)-\textbf{k}_T\cdot\textbf{p}_\perp}{zMM_h}h_1^\perp H_1^\perp\right]
\end{split}
\end{equation}
where: 
\begin{equation}
    \mathcal{C}[wfD] = \int \diff^2 \textbf{k}_T \diff^2 \textbf{p}_\perp \delta(\textbf{P}_T-z\textbf{k}_T-\textbf{p}_\perp)w f D
\end{equation}
and where $\hat{\textbf{h}} = \textbf{P}_T/|\textbf{P}_T|$. The $F_{UU,T}$ structure function is given by the convolution of the unpolarised TMD-PDF $f_1$ with the unpolarised TMD-FF $D_1$, while  $F_{UU}^{\cos2\phi_h}$ is written in terms of the Boer-Mulders TMD-PDFs $h_1^\perp$ and of the Collins FF $H_1^\perp$. The expression for $F_{UU}^{\cos\phi_h}$, in the Wandzura-Wilczek approximation \cite{Bastami:2018xqd}, contains a contribution proportional to the convolution of $h_1^\perp$ with $H_1^\perp$, together with a second term, proportional to the convolution of $f_1$ and $D_1$. The presence of this term was suggested long ago by Cahn \cite{Cahn:1978se,Cahn:1989yf}. The Cahn effect also contributes to the $F_{UU}^{\cos2\phi_h}$ structure function at twist-4: however, this is just one of the possible contributions arising at the same twist, generally not known \cite{Barone:2015ksa}. The convolutions $\mathcal{C}$ can be solved analytically e.g. assuming an exponential dependence of PDFs and FFs on the squared transverse momenta (the Gaussian Ansatz), according to which, for instance, an exponential trend is expected for $F_{UU,T}$:

\begin{equation}
    F_{UU,T} = \sum_{q}e_q^2 xf_1^q(x,Q^2) D_1^{h/q}(z,Q^2) ~ e^{-\frac{P_T^2}{\langle P_T^2\rangle}},
\end{equation}
where $\langle P_T^2 \rangle = z^2\langle k_T^2 \rangle + \langle p_\perp^2 \rangle$. A measurement of the $\phi_h$-integrated unpolarised SIDIS cross-section allows accessing the $P_T^2$-dependence of $F_{UU,T}$, while the features of the unpolarised TMDs can be accessed by measuring the amplitudes of the modulations in the azimuthal angle (azimuthal asymmetries) $A_{UU}^{\cos\phi_h}$, $A_{UU}^{\cos2\phi_h}$ and $A_{LU}^{\sin\phi_h}$ of the final-state hadrons.

\section{Data analysis}
In 2016 and 2017 the COMPASS Collaboration at CERN collected SIDIS data with a longitudinally polarised positive or negative muon beam scattering off a liquid hydrogen target. Part of these data has been analysed to investigate the unpolarised structure functions, through the measurement of the azimuthal asymmetries $A_{UU}^{\cos\phi_h}$, $A_{UU}^{\cos2\phi_h}$ and $A_{LU}^{\sin\phi_h}$ and of the distributions of the transverse momentum of the final state hadrons. The data used in this analysis were collected during the 2016 data taking and correspond to about 11\% of the total sample.

The selection of the DIS events has been performed requiring $Q^2>1$~(GeV/$c$)$^{2}$ and $W>5$~GeV/$c^2$. In addition, $x$ has been selected to be in the range $0.003<x<0.130$ (the upper limit being fixed by the reduced acceptance of the apparatus); events with $y<0.2$ or $y>0.9$ have been removed, to ensure a good event reconstruction and flat acceptance corrections and to limit the impact of radiative effects; the polar angle of the virtual photon $\theta_{\gamma^*}$ has been required to be smaller than 60~mrad. The hadrons have been selected to have $z>0.1$ and $P_{T}>0.1$~GeV/$c$. 

A non-negligible fraction of the selected hadron sample is constituted by the decay products of exclusive, diffractively produced vector mesons (mostly $\rho^0 \to \pi^+\pi^-$ and $\phi \to K^+K^-$), whose production mechanism can not be interpreted in the framework of the parton model. The decay products are concentrated at low $x$, low $P_T$ and high $z$; in addition, they show strong azimuthal modulations. Hence, their impact on the measured TMD observables has to be taken into account. In COMPASS, this has been done for both the  $P_T^2$-distributions \cite{COMPASS:2017mvk} and, more recently, for the azimuthal asymmetries \cite{COMPASS:2019lcm}. For the analysis of the 2016 data, the diffractive events have been selected (and discarded) by requiring $z_{tot}=z_{h^+}+z_{h^-}>0.95$; the remaining contribution of non-fully reconstructed pairs has been estimated and corrected for using the HEPGEN Monte Carlo \cite{Sandacz:2012at}, normalised to the data using the reconstructed pairs. 

The acceptance corrections have been evaluated using the LEPTO Monte Carlo \cite{Ingelman:1996mq} and applied in each kinematic bin, after correcting for the exclusive hadrons.

\section{Results}
\label{sec:another}
\subsection{Azimuthal asymmetries}

The three azimuthal asymmetries have been measured in bins of $x$, $z$ and $P_T$, both in a one-dimensional (1D) and in a three-dimensional (3D) approach. 
In the 1D case, the asymmetries have been inspected as a function of one variable, while integrating over the other two; in the 3D case, a simultaneous binning in $x$, $z$ and $P_T$ has been performed. In addition, the 1D approach has been applied also in four $Q^2$ bins. The azimuthal asymmetries have been obtained from a fit of the azimuthal distributions of the final-state hadrons after discarding (subtracting) the visible (non-visible) contribution of the decay hadrons from diffractively produced vector mesons and after correcting for acceptance. The fit procedure has been implemented independently for $\mu^+$ and $\mu^-$ data, and the corresponding results averaged after checking their compatibility.

\begin{figure}[h]
\centering
\includegraphics[width=0.45\textwidth]{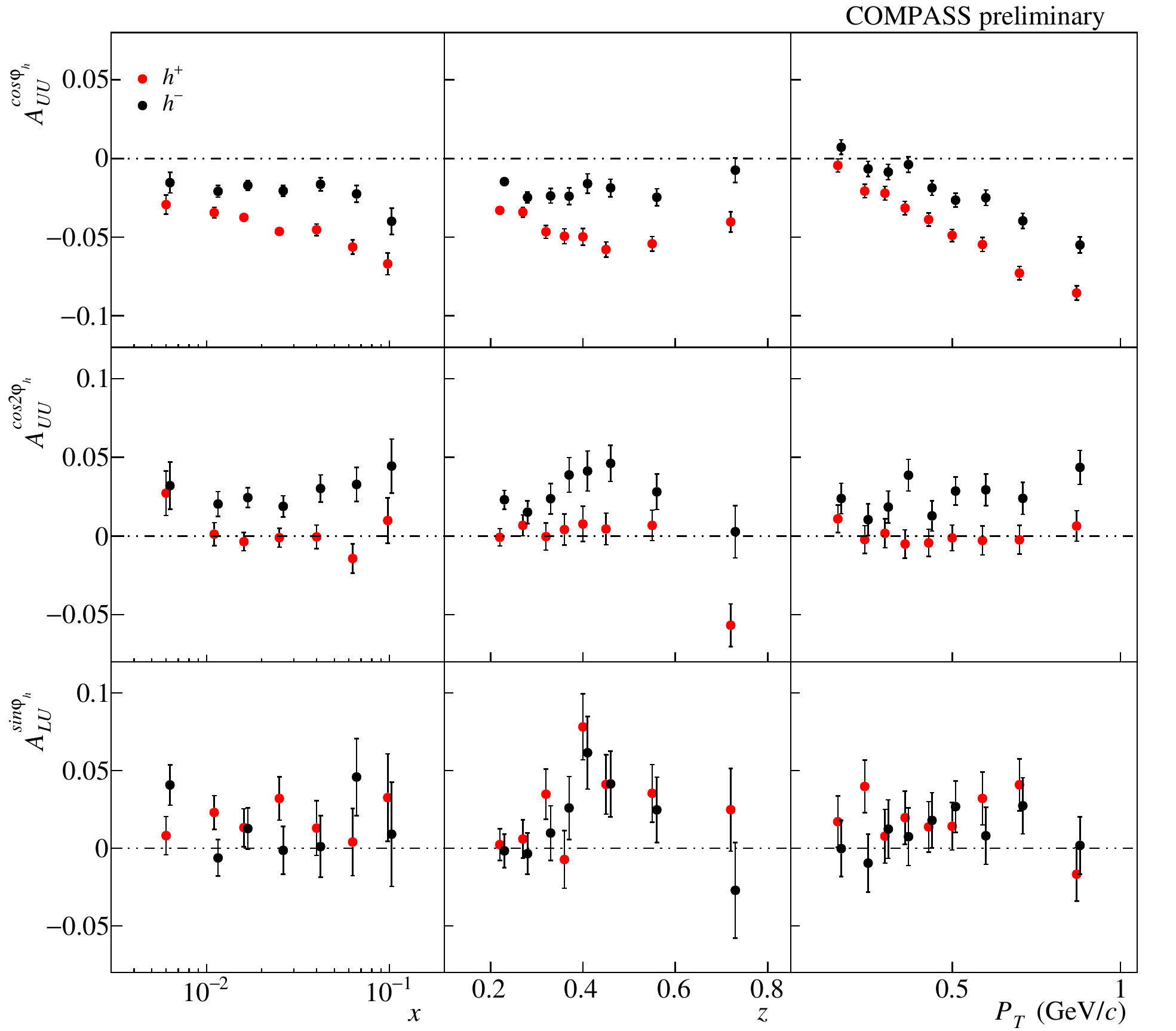}
\includegraphics[width=0.50\textwidth]{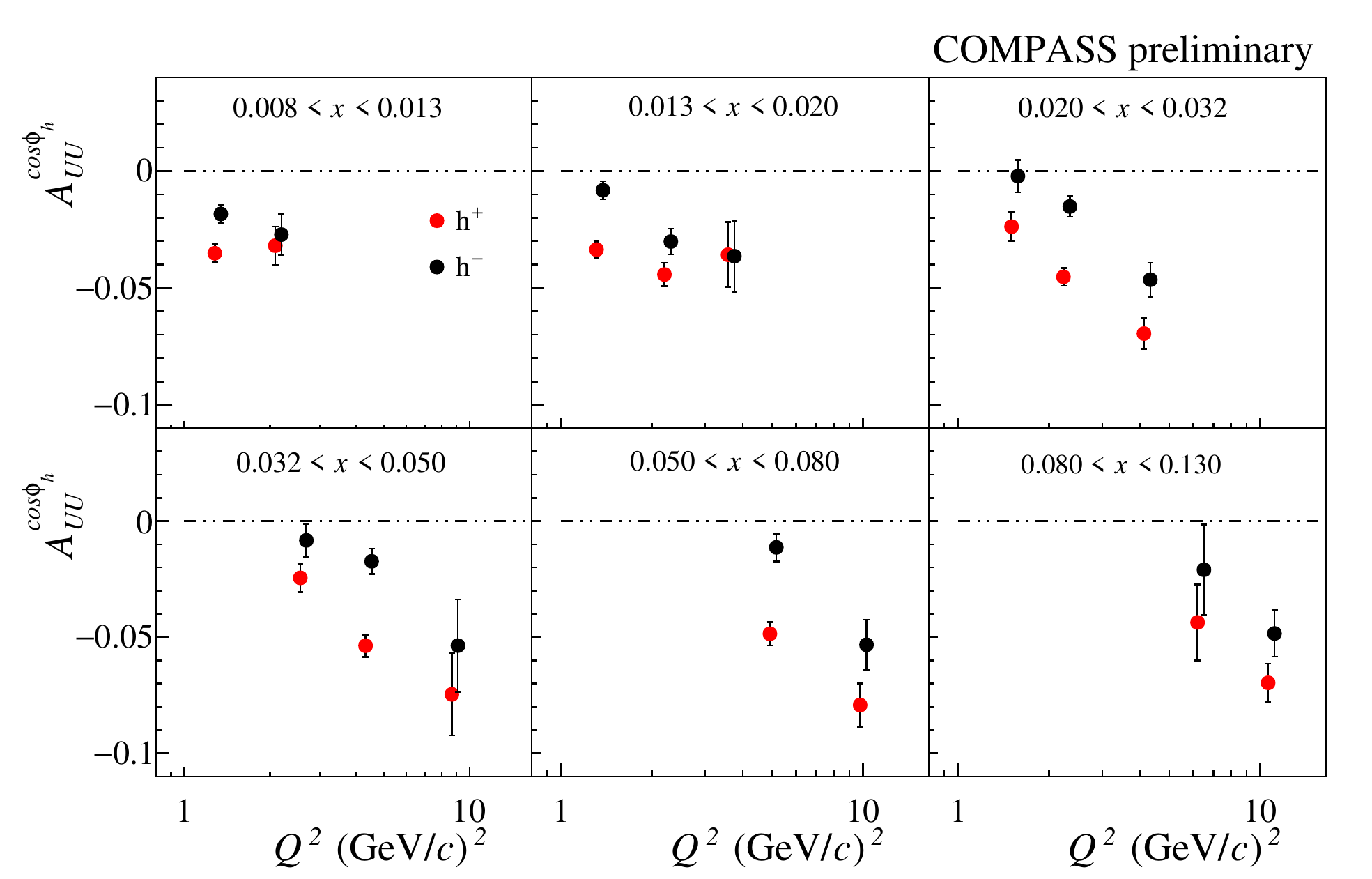}
\caption{Left: Azimuthal asymmetries $A_{UU}^{\cos\phi_h}$ (top row), $A_{UU}^{\cos2\phi_h}$ (middle row) and $A_{LU}^{\sin\phi_h}$ (bottom row) for positive (red) and negative hadrons (black). Right: $Q^2$-dependence of $A_{UU}^{\cos\phi_h}$ in bins of $x$ for positive (red) and negative hadrons (black).}
\label{fig:aares1Dstd}
\end{figure}

\begin{figure}[h]
\centering
\includegraphics[width=0.6\textwidth]{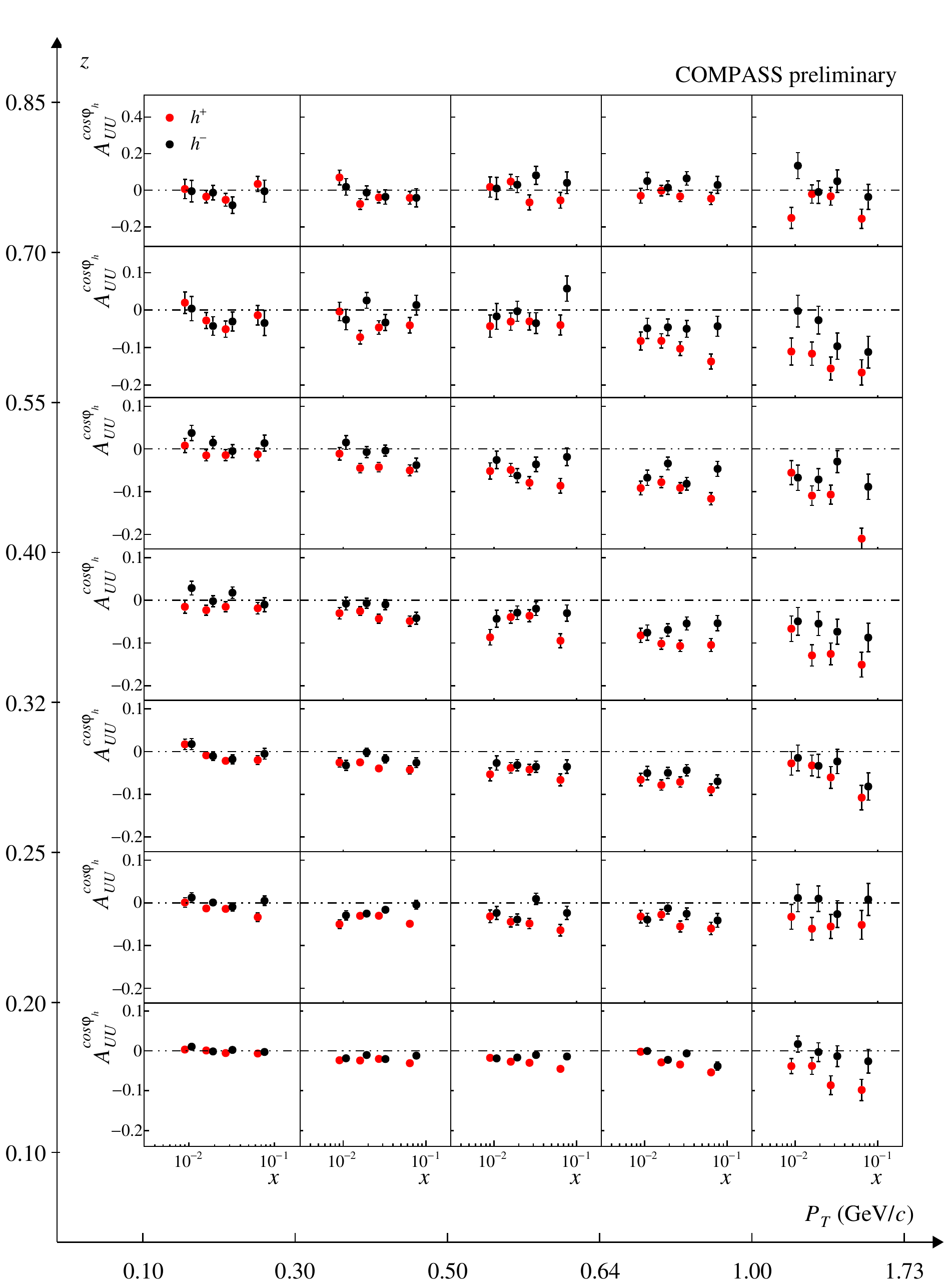}
\caption{$A_{UU}^{\cos\phi_h}$ asymmetry for positive (red) and negative hadrons (black), as a function of $x$ and in bins of $z$ (vertical axis) and $P_T$ (horizontal axis).}
\label{fig:aares3D0}
\end{figure}

The 1D results are shown in Fig.~\ref{fig:aares1Dstd} (left). The $A_{UU}^{\cos\phi_h}$ asymmetries are large, clearly different from zero, with a linear trend in $P_T$ and an interesting difference between positive and negative hadrons, not expected from a flavor-independent Cahn mechanism but already observed in a previous COMPASS measurement on deuteron \cite{COMPASS:2014kcy}. The $A_{UU}^{\cos2\phi_h}$ asymmetry is generally smaller, being compatible with zero for positive hadrons and different from zero for negative hadrons. The $A_{LU}^{\sin\phi_h}$ asymmetry shows larger uncertainties due to the kinematic corrections. The $A_{UU}^{\cos\phi_h}$ asymmetries also shows a remarkable dependence on $Q^2$, as shown in Fig.~\ref{fig:aares1Dstd} (right). The 3D results for $A_{UU}^{\cos\phi_h}$ are shown in Fig.~\ref{fig:aares3D0}: this asymmetry, generally of negative sign as expected from the Cahn effect, is characterised by a smooth dependence on $x$ and $P_T$, becoming compatible with zero at large $z$. The shown uncertainties are statistical only; the systematic uncertainty has been estimated to be $\sigma_{syst} \sim ~\sigma_{stat}$ in the 1D case. 

\subsection{$P_T^2$—distributions}
\begin{figure}[h!]
\centering
\includegraphics[width=0.6\textwidth]{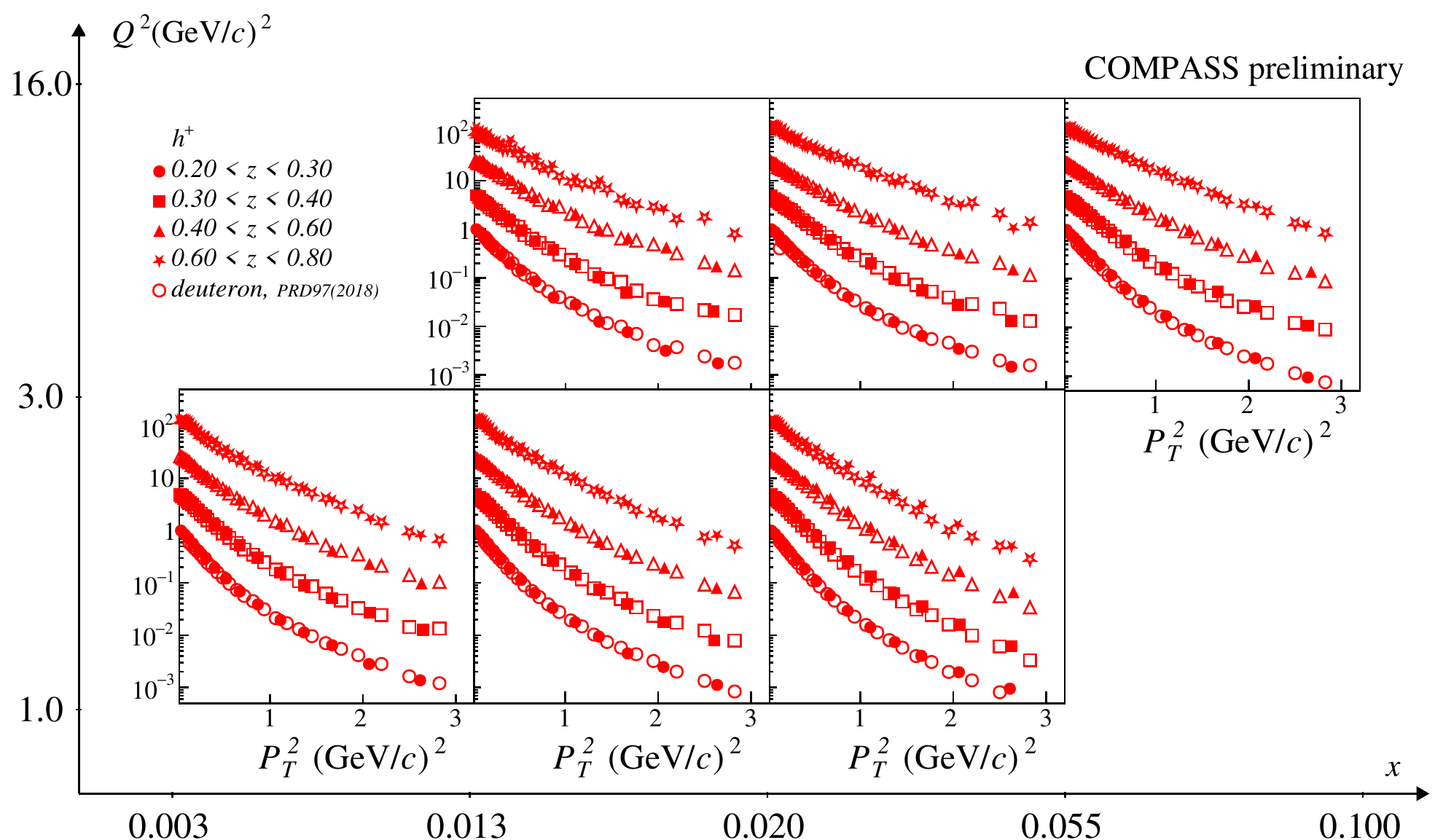}
\caption{$P_T^2$—distributions for positive hadrons in four $z$ bins and in bins of $x$ and $Q^2$, for the new preliminary results on proton (full markers) and for the published deuteron results \cite{COMPASS:2017mvk}.}
\label{fig:ptds}
\end{figure}

\begin{figure}[h!]
\centering
\includegraphics[width=0.6\textwidth]{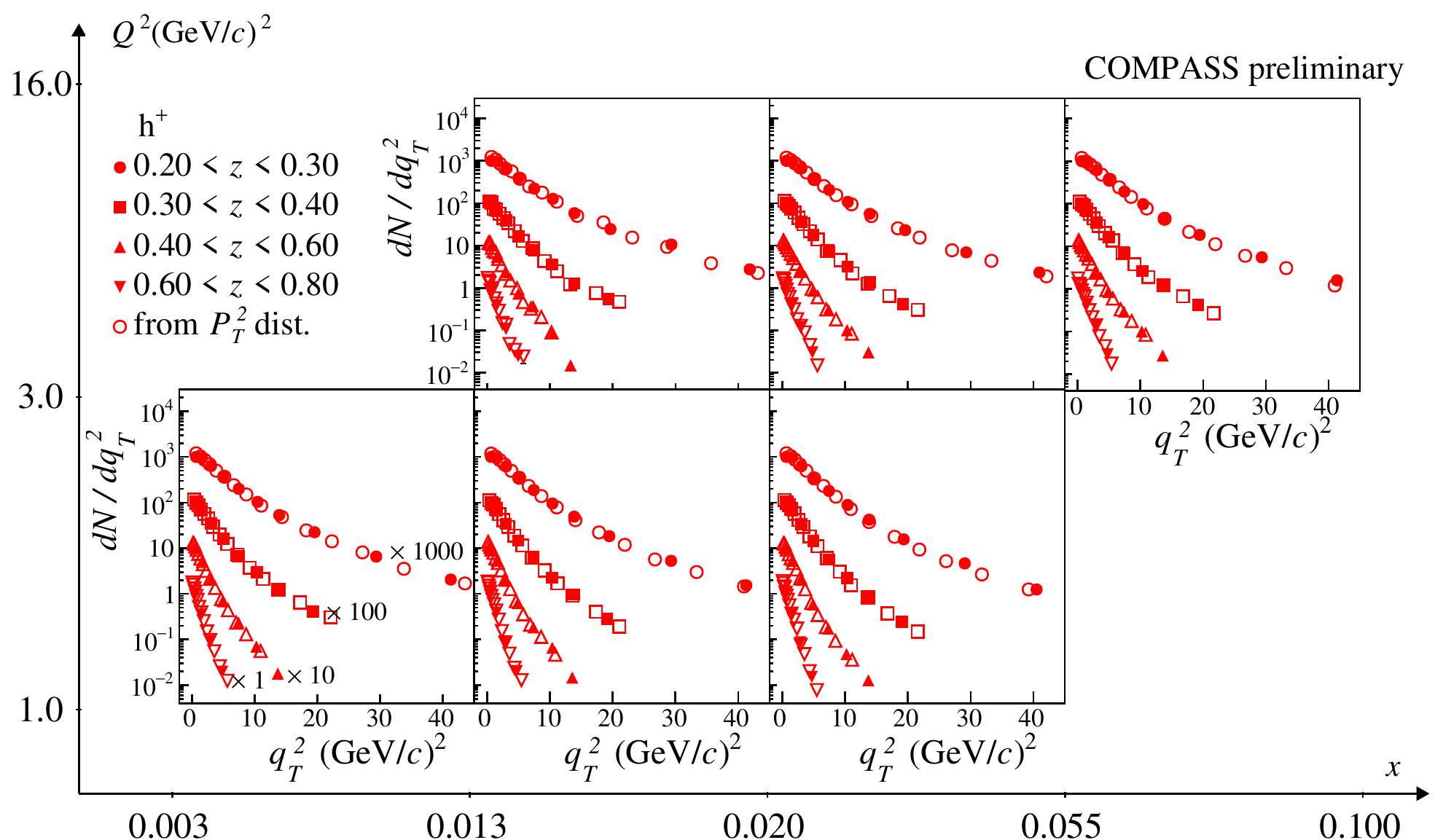}
\caption{$q_T^2$—distributions for positive hadrons in four $z$ bins and in bins of $x$ and $Q^2$, as obtained with a direct extraction (full markers) and (open markers) from a conversion of the $P_T^2$-distributions in Fig.~\ref{fig:ptds}.}
\label{fig:qtds}
\end{figure}

The transverse-momentum distributions have been produced separately for $\mu^+, \mu^-, h^+, h^-$ after the exclusive hadrons and acceptance corrections; they have been normalised to the value in the first bin in $P_T^2$ and averaged over the beam charge. Figure~\ref{fig:ptds} shows the distributions for positive hadrons up to $P_T^2 = 3$~(GeV/$c$)$^2$ in the four $z$ bins (full markers), and in bins of $x$ and $Q^2$, with a further arbitrary normalization introduced for a better readability. At low $P_T^2$, the trend is exponential, as expected, while a second component at higher $P_T^2$ can also be seen. While the distributions do not exhibit a significant $x$-dependence, a mild dependence on $Q^2$ can instead be observed, with a larger $\langle P_T^2\rangle$ at higher $Q^2$. The dependence on $z$ is remarkable. No significant difference between the shape of positive and negative hadrons distributions has been found. These shapes can be used to get independent information on $\langle k_T^2 \rangle$ through the relation $\langle P_T^2 \rangle = z^2\langle k_T^2 \rangle + \langle p_\perp^2 \rangle$. The shown uncertainties are statistical only; the systematic uncertainty has been estimated to be $\sigma_{syst} = 0.3~\sigma_{stat}$. The new results are compared in the same plot to the published COMPASS results on deuteron \cite{COMPASS:2017mvk} (open markers): the two sets of results are in qualitative agreement, particularly at small $z$. In addition, the distributions of $q_T^2=P_T^2/z^2$ have been produced from the same samples: they are shown for positive hadrons in Fig.~\ref{fig:qtds} (full markers) where they are compared with the distributions obtained from the conversion of the $P_T^2$-distributions, done using $\diff q_T \approx \diff P_T /z$ (empty markers). A good agreement can be observed between the two extractions both in size and shape.

\section{Conclusion}
In 2016 and 2017, the COMPASS Collaboration at CERN collected SIDIS data using an unpolarised proton target. Preliminary results for the azimuthal asymmetries and the transverse momentum distributions, which give important information on the TMD structure of the nucleon, have been produced from part of the data, taking into account the non-negligible contribution of the hadrons produced in the decay of diffractively produced vector mesons. The azimuthal asymmetries shown here exhibit clear and strong kinematic dependences, while the trend of the $P_T^2$-distributions is smooth, as already observed in previous COMPASS measurements. The inclusion of more data in the analysis and the use of refined Monte Carlo simulations will increase the statistical precision and reduce the systematic uncertainties, also allowing for a direct measurement of the transverse-momentum-dependent unpolarised SIDIS cross-section, to be done in the near future. 

% TODO:
% Provide your bibliography here. You have two options:

% FIRST OPTION - write your entries here directly, following the example below, including Author(s), Title, Journal Ref. with year in parentheses at the end, followed by the DOI number.
%\begin{thebibliography}{99}
%\bibitem{Bacchetta:2006tn} A. Bacchetta, M. Diehl, K. Goeke, A. Metz, P. J. Mulders and M. Schlegel, {\it Semi-inclusive deep inelastic scattering at small transverse momentum}, JHEP {\bf 02}, 093 (2007), \doi{10.1088/1126-6708/2007/02/093}, \url{http://arxiv.org/abs/hep-ph/0611265}.
%\bibitem{arXiv:1108.2700} P. Ginsparg, {\it It was twenty years ago today... }, \url{http://arxiv.org/abs/1108.2700}.
%\end{thebibliography}

%\bibliographystyle{SciPost_bibstyle} % Include this style file here only if you are not using our template
\bibliography{bibliography.bib}

\begin{thebibliography}{10}
\providecommand{\url}[1]{\texttt{#1}}
\providecommand{\urlprefix}{URL }
\expandafter\ifx\csname urlstyle\endcsname\relax
  \providecommand{\doi}[1]{doi:\discretionary{}{}{}#1}\else
  \providecommand{\doi}{doi:\discretionary{}{}{}\begingroup
  \urlstyle{rm}\Url}\fi
\providecommand{\eprint}[2][]{\url{#2}}

\bibitem{Bacchetta:2006tn}
A.~Bacchetta, M.~Diehl, K.~Goeke, A.~Metz, P.~J. Mulders and M.~Schlegel,
\newblock \emph{{Semi-inclusive deep inelastic scattering at small transverse
  momentum}},
\newblock JHEP \textbf{02}, 093 (2007),
\newblock \doi{10.1088/1126-6708/2007/02/093},
\newblock \eprint{https://arxiv.org/abs/hep-ph/0611265}.

\bibitem{Collins:2011zzd}
J.~Collins,
\newblock \emph{{Foundations of perturbative QCD}}, vol.~32,
\newblock Cambridge University Press,
\newblock ISBN 978-1-107-64525-7 (2013).

\bibitem{Bastami:2018xqd}
S.~Bastami \emph{et~al.},
\newblock \emph{{Semi-Inclusive Deep Inelastic Scattering in
  Wandzura-Wilczek-type approximation}},
\newblock JHEP \textbf{06}, 007 (2019),
\newblock \doi{10.1007/JHEP06(2019)007},
\newblock \eprint{https://arxiv.org/abs/1807.10606}.

\bibitem{Cahn:1978se}
R.~N. Cahn,
\newblock \emph{{Azimuthal Dependence in Leptoproduction: A Simple Parton Model
  Calculation}},
\newblock Phys. Lett. B \textbf{78}, 269 (1978),
\newblock \doi{10.1016/0370-2693(78)90020-5}.

\bibitem{Cahn:1989yf}
R.~Cahn,
\newblock \emph{{Critique of Parton Model Calculations of Azimuthal Dependence
  in Leptoproduction}},
\newblock Phys. Rev. D \textbf{40}, 3107 (1989),
\newblock \doi{10.1103/PhysRevD.40.3107}.

\bibitem{Barone:2015ksa}
V.~Barone, M.~Boglione, J.~O. Gonzalez~Hernandez and S.~Melis,
\newblock \emph{{Phenomenological analysis of azimuthal asymmetries in
  unpolarized semi-inclusive deep inelastic scattering}},
\newblock Phys. Rev. D \textbf{91}(7), 074019 (2015),
\newblock \doi{10.1103/PhysRevD.91.074019},
\newblock \eprint{https://arxiv.org/abs/1502.04214}.

\bibitem{COMPASS:2017mvk}
M.~Aghasyan \emph{et~al.},
\newblock \emph{{Transverse-momentum-dependent Multiplicities of Charged
  Hadrons in Muon-Deuteron Deep Inelastic Scattering}},
\newblock Phys. Rev. D \textbf{97}(3), 032006 (2018),
\newblock \doi{10.1103/PhysRevD.97.032006},
\newblock \eprint{https://arxiv.org/abs/1709.07374}.

\bibitem{COMPASS:2019lcm}
J.~Agarwala \emph{et~al.},
\newblock \emph{{Contribution of exclusive diffractive processes to the
  measured azimuthal asymmetries in SIDIS}},
\newblock Nucl. Phys. B \textbf{956}, 115039 (2020),
\newblock \doi{10.1016/j.nuclphysb.2020.115039},
\newblock \eprint{https://arxiv.org/abs/1912.10322}.

\bibitem{Sandacz:2012at}
A.~Sandacz and P.~Sznajder,
\newblock \emph{{HEPGEN - generator for hard exclusive leptoproduction}}
  (2012),
\newblock \eprint{https://arxiv.org/abs/1207.0333}.

\bibitem{Ingelman:1996mq}
G.~Ingelman, A.~Edin and J.~Rathsman,
\newblock \emph{{LEPTO 6.5: A Monte Carlo generator for deep inelastic lepton -
  nucleon scattering}},
\newblock Comput. Phys. Commun. \textbf{101}, 108 (1997),
\newblock \doi{10.1016/S0010-4655(96)00157-9},
\newblock \eprint{https://arxiv.org/abs/hep-ph/9605286}.

\bibitem{COMPASS:2014kcy}
C.~Adolph \emph{et~al.},
\newblock \emph{{Measurement of azimuthal hadron asymmetries in semi-inclusive
  deep inelastic scattering off unpolarised nucleons}},
\newblock Nucl. Phys. B \textbf{886}, 1046 (2014),
\newblock \doi{10.1016/j.nuclphysb.2014.07.019},
\newblock \eprint{https://arxiv.org/abs/1401.6284}.

\end{thebibliography}

\nolinenumbers

\end{document}